\documentclass[preprint,12pt,compress]{elsarticle}

\makeatletter
\def\ps@pprintTitle{}
\makeatother

\usepackage{hyperref}
\usepackage{url}
\usepackage{booktabs}
\usepackage{mathrsfs}
\usepackage{array}
\usepackage{graphicx}
\usepackage{latexsym}
\usepackage{amsmath}
\usepackage{amsthm}
\usepackage{color}
\usepackage{xcolor}
\usepackage{amsfonts}
\usepackage{subfigure}
\usepackage{dsfont}
\usepackage{epstopdf}
\usepackage{threeparttable}
\usepackage{multirow}
\usepackage{enumerate}
\usepackage{epsfig}
\usepackage{bm}
\usepackage{algorithmic}
\usepackage{textcomp}
\usepackage{graphicx}
\usepackage{latexsym}
\usepackage{amsmath}
\usepackage{amsthm}
\usepackage{makecell,rotating,multirow,diagbox}
\usepackage{amsfonts}
\usepackage{subfigure}
\usepackage{dsfont}
\usepackage{epstopdf}
\usepackage{threeparttable}
\usepackage{multirow}
\usepackage{enumerate}
\usepackage{epsfig}
\usepackage{bm}
\usepackage{algorithmic}
\usepackage{textcomp}
\usepackage{xcolor}
\usepackage{caption}
\usepackage{booktabs}

\usepackage{array, caption, threeparttable}

\usepackage[font=small,labelfont=bf,labelsep=none]{caption}

\usepackage{amssymb}

\usepackage{tabu}
\usepackage{makecell}

\usepackage{amssymb}

\journal{Physical Communication}

\begin{document}

\begin{frontmatter}



\title{Amplitude Prediction from Uplink to Downlink CSI against Receiver Distortion in FDD Systems}


\author{Chaojin Qing\textsuperscript{a,}$^{\ast}$, Zilong Wang\textsuperscript{a}, Qing Ye\textsuperscript{a}, Wenhui Liu\textsuperscript{a}, and Linsi He\textsuperscript{a}}

\affiliation{organization={School of Electrical Engineering and Electronic Information, Xihua University},
            city={Chengdu},
            postcode={610039}, 
            country={China}}

\begin{abstract}

	In frequency division duplex (FDD) massive multiple-input multiple-output (mMIMO) systems, the reciprocity mismatch caused by receiver distortion seriously degrades the amplitude prediction performance of channel state information (CSI). To tackle this issue, from the perspective of distortion suppression and reciprocity calibration, a lightweight neural network-based amplitude prediction method is proposed in this paper. Specifically, with the receiver distortion at the base station (BS), conventional methods are employed to extract the amplitude feature of uplink CSI. Then, learning along the direction of the uplink wireless propagation channel, a dedicated and lightweight distortion-learning network (Dist-LeaNet) is designed to restrain the receiver distortion and calibrate the amplitude reciprocity between the uplink and downlink CSI. Subsequently, by cascading, a single hidden layer-based amplitude-prediction network (Amp-PreNet) is developed to accomplish amplitude prediction of downlink CSI based on the strong amplitude reciprocity. Simulation results show that, considering the receiver distortion in FDD systems, the proposed scheme effectively improves the amplitude prediction accuracy of downlink CSI while reducing the transmission and processing delay.

\end{abstract}



\begin{keyword}
	 CSI feedback \sep massive MIMO \sep amplitude prediction \sep receiver distortion \sep lightweight network



\end{keyword}

\end{frontmatter}


\section{Introduction}
\label{section:1}
As one of the key techniques in the fifth generation (5G) communications, the massive multiple-input multiple-output (mMIMO) has shown great prospects in providing high spectrum and energy efficiency \cite{I1,A1,R1-1}. In frequency division duplex (FDD) systems, the downlink channel state information (CSI) estimated by user equipment (UE) usually needs to be fed back to the base station (BS) \cite{A3}. However, due to the large number of antennas, the CSI feedback overhead in mMIMO systems increases sharply, which results in large transmission and processing delay, energy consumption, and transmission resource occupation, etc \cite{A4}. Especially, in high-speed scenarios, the transmission and processing delay may cause the downlink CSI obtained at the BS to be outdated \cite{A5}. Therefore, it is crucial to reduce the feedback overhead and processing delay in FDD mMIMO systems.

Recently, some studies have shown that there is a strong amplitude correlation (or reciprocity \footnote{Note that, due to the small enough difference in frequency-independent parameters between the uplink CSI and downlink CSI in FDD systems, we assume the reciprocity in this paper.}) between the uplink CSI and downlink CSI in FDD systems \cite{II3, I2}. Hence, it becomes popular to use deep learning (DL) to directly predict the downlink CSI from the uplink CSI to reduce/eliminate the feedback process \cite{I3,I5,I6}. In \cite{I3}, considering the position-to-channel mapping is bijective, a sparse complex-valued neural network (SCNet) is proposed to approximate the uplink-to-downlink mapping function. In \cite{I5}, according to the spatial correlation in time-varying scenarios, a convolutional neural network (CNN)-based downlink channel prediction method is investigated. Under the premise of considering the channel time invariance, an attention-based deep learning network is proposed in \cite{I6} to directly predict the downlink CSI from the uplink CSI.

In \cite{II3,I2,I3,I5,I6}, the feedback and prediction methods are mainly based on the reciprocity between the uplink and downlink CSIs. However, this reciprocity is vulnerable in practical systems. This is because the overall channel consists of not only the wireless propagation channel, but also the radio frequency (RF) front-end, e.g., analog-to-digital converters (ADCs), filters and low-noise amplifiers (LNAs), etc \cite{I7,A6}.
Although uplink and downlink wireless propagation channels may be reciprocal, the hardware imperfection (HI) of RF front-end inevitably introduces nonlinear distortion into the amplitude and phase of the transmitted and received signals \cite{R2-3}.
This nonlinear distortion causes the amplitude mismatch and phase mismatch, thereby resulting in the reciprocity mismatch between the uplink and downlink CSIs.

Due to the significant impact of reciprocity mismatch on system performance, reciprocity calibration is crucial for communication systems \cite{I7}.
In both regular MIMO and mMIMO systems, the conventional reciprocity calibration method is based on the dedicated hardware circuits \cite{I9}, which increases the energy consumption and hardware cost that comes from RF chains required to support a number of antennas \cite{I8}.
Besides, the existing reciprocity calibration is usually investigated in time division duplex (TDD) systems. In contrast, there is limited literature addressing the issue of reciprocity calibration in FDD systems when utilizing the reciprocity.
That is, in practical FDD mMIMO systems, the reciprocity between uplink and downlink channels is destroyed by the nonlinear distortion due to the difference between uplink and downlink hardware, making the reciprocity-based prediction results inaccurate, and even resulting in the prediction process impossible to achieve.

Therefore, it is vital to consider the nonlinear distortion before utilizing reciprocity and the reciprocity calibration is also essential for the reciprocity-based amplitude prediction in FDD systems. To suppress the impact of distortion on reciprocity and calibrate the reciprocity, this paper proposes an amplitude prediction scheme against receiver distortion. To the best of our knowledge, the amplitude reciprocity-based CSI prediction by considering distortion has not been investigated in FDD systems. The main contributions of this paper are summarized as follows:
\begin{itemize}
	\item  We propose a more practical scenario which considers the distortion before utilizing the reciprocity in FDD systems. Specifically, we take the receiver distortion at BS as an example to illustrate that reciprocity will be affected by distortion, which is a valuable reference for both UE and BS.

	\item We design a network architecture cascading distortion learning and amplitude prediction to improve the practicality and accuracy of amplitude prediction. Specifically, learning along the direction of the uplink wireless propagation channel, a distortion-learning network (Dist-LeaNet) is designed to restrain the receiver distortion and calibrate the amplitude reciprocity between the uplink and downlink CSI.
	Subsequently, based on the channel amplitude reciprocity, an amplitude-prediction network (Amp-PreNet) is developed to predict the amplitude of downlink CSI directly at the BS, thus avoiding the overhead and transmission delay caused by feedback.
	
	\item We construct a lightweight learning and prediction network architecture to reduce the processing delay and computational complexity for the BS receiver. Due to that the nonlinear distortion varies slowly compared with the wireless propagation channel, the features of distortion are easy to capture. Hence, Dist-LeaNet is constructed with a lightweight network architecture. With the assistance of Dist-LeaNet, Amp-PreNet is also constructed with lightweight network architecture based on the strong amplitude correlation.

\end{itemize}

The rest of this paper is organized as follows. In Section \ref{section:2}, we introduce the system model. Then, the amplitude prediction scheme against receiver distortion is presented in Section \ref{section:3} and followed by numerical results in Section \ref{section:4}. Finally, Section \ref{section:5} concludes our work.

\textit{Notation}: Boldface upper case and lower case letters denote matrix and vector respectively. ${\cal N}\left( {\mu ,{\sigma ^2}} \right)$ stands for normal distribution with mean $\mu $ and variance ${{\sigma ^2}}$; ${\cal U}\left( {a,b} \right)$ stands for uniform distribution on the interval $\left( {a,b} \right)$; $|\cdot|$ denotes the operation of taking the modulus of a complex value; ${\left(\cdot \right)^T}$ denotes transpose; $E[\cdot]$ represents the expectation operation; ${\left\|  \cdot  \right\|}$ is the Euclidean norm.

\section{System Model}
\label{section:2}

\begin{figure}[t]
	\centering
	\includegraphics[width=5.5in]{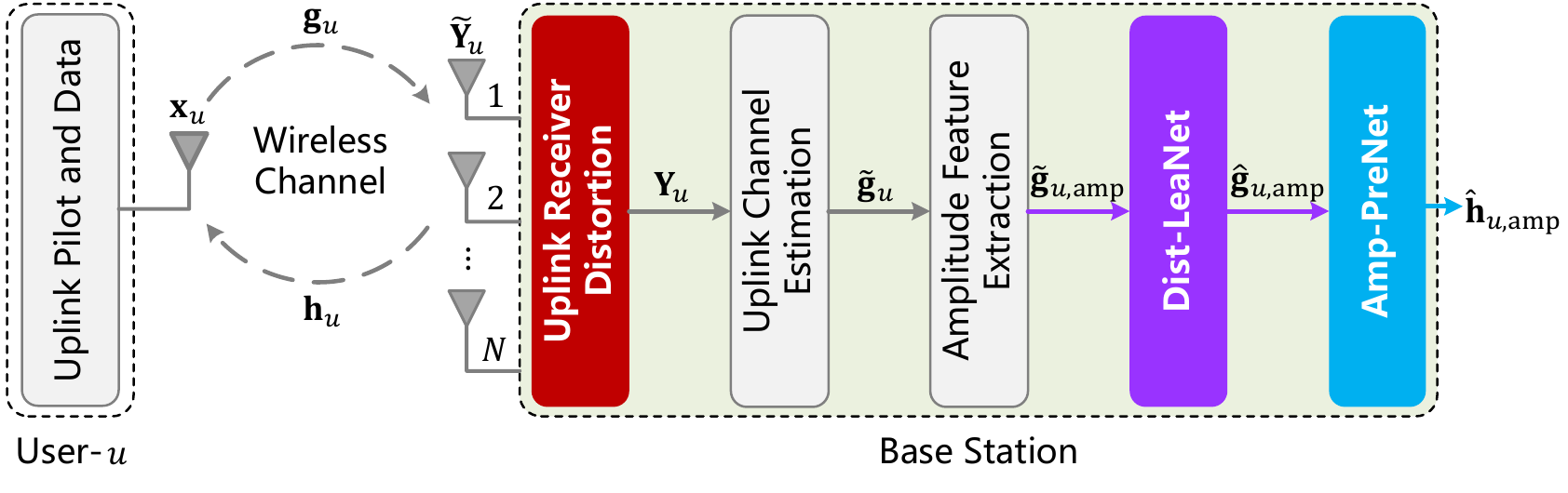}
	\caption{System model.\label{fig1}}
\end{figure}

The system model is given in Fig.~\ref{fig1}, in which an FDD massive MIMO system that consists of a BS with $N$ antennas and $U$ single-antenna users in speed $v$ is considered. At the BS, the received uplink signal from user-$u$, denoted as ${\widetilde {\bf{Y}}_u} \in \mathbb{C}^{N\times N}$, is given by
\begin{equation}
	\label{equ:received_signal_y}
	{\widetilde {\bf{Y}}_u} = {{\bf{g}}_u}{{\bf{x}}_u} + {{\bf{N}}_u},
\end{equation}
where ${{\bf{g}}_u} \in {{\Bbb C}^{N \times 1}}$ denotes the uplink channel (i.e., uplink CSI) from the user-$u$ to the BS in the angular domain, ${{\bf{x}}_u} \in \mathbb{C}^{1\times N}$ stands $N$-length uplink pilot and data of user-$u$, and ${{\bf{N}}_u} \in {{\Bbb C}^{N \times N}}$ is the circularly symmetric complex Gaussian (CSCG) noise with zero-mean and variance ${\sigma_{u}^2}$.
With the antenna diversity \cite{RPAbook}, the uplink pilot and data of each UE is received by $N$ BS antennas to form the $N \times N$ complex signal ${\widetilde {\bf{Y}}_u}$.
From \cite{III10}, there exists a correlation between the uplink and downlink channels due to the shared common physical paths and similar spatial propagation characteristics.
For example, the downlink CSI is constructed by utilizing frequency-independent parameters between the uplink and downlink channels in the angular domain \cite{RPA1}.
Therefore, from \cite{II2,R2-1}, the downlink CSI of user-$u$ (i.e., ${\bf{h}}_u \in \mathbb{C}^{N\times 1}$) can be recovered from the uplink CSI ${{\bf{g}}_u}$. However, the inevitable uplink distortion (e.g., caused by imperfect hardware) makes this processing difficult.
By denoting the mapping function of equivalent distortion at the BS as ${f_{{\text{R-dis}}}}\left(  \cdot  \right)$, the distorted signal of ${\widetilde {\bf{Y}}_u}$, denoted by ${ {\bf{Y}}_u} \in \mathbb{C}^{N\times N}$, is expressed as
\begin{equation}
	\label{equ:distortion_received_signal_y}
	{ {\bf{Y}}_u} \triangleq {f_{{\text{R-dis}}}}\left( {{\widetilde{\bf{Y}}_u}} \right).
\end{equation}
Then, the uplink channel ${\bf{g}}_u$ is estimated according to ${ {\bf{Y}}_u}$ and the uplink pilot in ${{\bf{x}}_u}$. By denoting the estimated ${\bf{g}}_u$ as $\widetilde{\bf{g}}_u$ ($\widetilde{\bf{g}}_u \in \mathbb{C}^ {N \times 1}$), our work aims to utilize the strong amplitude correlation to predict the amplitude of downlink CSI $\bf{h}_u$ from $\widetilde{\bf{g}}_u$ \cite{II3, I2} in this paper. First, the amplitude feature of $\widetilde{\bf{g}}_u$ is extracted.
Subsequently, we build two dedicated networks, Dist-LeaNet and Amp-PreNet, to restrain the distortion of receiver and enhance the prediction accuracy of downlink CSI amplitude, respectively. The details are described in Section \ref{section:3}.
However, the phase information exhibits unique importance due to its frequency-dependent nature, which is directly fed back to the BS \cite{I2}.

\section{Amplitude Prediction Scheme against Receiver Distortion}
\label{section:3}
To effectively utilize the reciprocity in practical scenarios, we present the proposed amplitude prediction scheme against receiver distortion in this section. In Section \ref{section:3.1}, the model of uplink receiver distortion is presented. With the uplink receiver distortion at the BS, we develop Dist-LeaNet and Amp-PreNet to restrain the distortion and predict the downlink CSI amplitude, respectively. Both Dist-LeaNet and Amp-PreNet are elaborated in Section \ref{section:3.2}.

\subsection{Uplink Receiver Distortion}
\label{section:3.1}

In the uplink communication, nonlinear distortion is inevitably encountered \cite{A7}, e.g., the distortion of user's power amplifiers (PAs), the distortion of BS's LNAs and ADCs, etc \cite{III1}. We mainly take the receiver distortion of BS as an example to represent the uplink distortion, which has reference value for both UE and BS.

Specifically, the nonlinear distortion varies slowly compared with the wireless propagation channel \cite{A8}.
Therefore, to further represent the distortion function ${f_{{\text{R-dis}}}}\left(  \cdot  \right)$, the receiver distortion at the BS is denoted as ${{\bf{D}}_{{\text{R-BS}}}} \in {{\Bbb C}^{N \times N}}$, wherein its diagonal elements represent the amplitude and phase distortion of each hardware at different antennas, and the off-diagonal elements correspond to the crosstalk and mutual coupling effect between different antennas \cite{III1}. The proper hardware circuit design can ensure the nearly-zero crosstalk, and the antenna mutual coupling effect is often ignored \cite{III2}. Therefore, the receiver distortion matrix can be regarded to be diagonal, which is expressed as \cite{III3}
\begin{equation}
	\label{equ:distortion_R_BS}
	{{\bf{D}}_{{\text{R-BS}}}} = {\text{diag}}\left( {{r_{1{\text{,BS}}}}, \cdots ,{r_{n{\text{,BS}}}}, \cdots ,{r_{N{\text{,BS}}}}} \right),
\end{equation}
where ${r_{n,{\text{BS}}}} = \left| {{r_{n,{\text{BS}}}}} \right|{e^{j\phi _{n,{\text{BS}}}^{\text{r}}}}$ $\left( {n = 1,2, \cdots ,N} \right)$.
According to \cite{III3}, the amplitudes of the distortion obey log-normal distribution, and the phases of the distortion obey uniform distribution, i.e.,
\[{\rm{ln}}\left| {{r_{n,{\rm{BS}}}}} \right| \sim {\cal N}\left( {0,\delta _{{\rm{r,BS}}}^2} \right),\phi _{n,{\rm{BS}}}^{\rm{r}} \sim {\cal U}\left[ { - {\theta _{{\rm{r,BS}}}},{\theta _{{\rm{r,BS}}}}} \right].\]

\subsection{Dist-LeaNet and Amp-PreNet}
\label{section:3.2}
In order to restrain the receiver distortion and obtain amplitude feature of the uplink wireless propagation channel, we construct the lightweight and effective Dist-LeaNet, which is supposed to be considered when channel reciprocity is involved. Then, a recovered uplink CSI amplitude feature ${\widehat {\bf{g}}_{u,{\text{amp}}}}$, is learned from Dist-LeaNet. Subsequently, to predict the downlink CSI amplitude feature, we design the lightweight Amp-PreNet, which utilizes the amplitude correlation of CSI in the angular domain \cite{I2}. The corresponding network design, training and deployment are as follows.

\subsubsection{Network Design}
\label{section:3.2.1}
According to \cite{III5}, choosing the appropriate number of layers and hidden neurons is still a challenge in the neural network (NN).
That is, for a specific network design, there is currently no established theoretical guidance on the optimal number of layers or the number of neurons to be included at each layer. Typically, complex hyper-parameter tuning is necessary.
Based on plenty of experimental results, we design the lightweight Dist-LeaNet and Amp-PreNet, both of which are single hidden-layer NN.
Specifically, considering the trade-off between performance and complexity, we train the network with different number of layers and neurons.
After verifying the performance of the trained network, we select a suitable lightweight network architecture to reduce the computational complexity while improve the prediction performance compared with \cite{II3}.
The network architectures of Dist-LeaNet and Amp-PreNet are summarized in Table~\ref{table_I}, and the detailed descriptions are given as follows.

\begin{table}[]
	
	\renewcommand{\arraystretch}{1.5}
	\caption{Architecture of Dist-LeaNet and Amp-PreNet.}
	\label{table_I}
	\centering
	\scalebox{0.64}{
		\begin{tabu}{@{}c|c|c|c|c|c|c@{}}
			\tabucline[0.8pt]{-}
			\multirow{2}{*}{Layer} & \multicolumn{2}{c|}{Input} & \multicolumn{2}{c|}{Hidden} & \multicolumn{2}{c}{Output} \\ \cline{2-7}
			& \multicolumn{1}{c|}{Dist-LeaNet} & \multicolumn{1}{c|}{Amp-PreNet} & \multicolumn{1}{c|}{Dist-LeaNet} & \multicolumn{1}{c|}{Amp-PreNet} & \multicolumn{1}{c|}{Dist-LeaNet} & \multicolumn{1}{c}{Amp-PreNet} \\ \tabucline[0.6pt]{-}
			Batch normalization &$\surd$ &$\times$ & $\times$ & $\times$ &$\times$ &$\times$ \\ \hline
			Neuron number       & $N$    & $N$     & $2N$  & $2N$    & $N$     & $N$    \\ \hline
			Activation function & None & None & Linear & LReLU & Linear & Linear  \\ \tabucline[0.8pt]{-}
		\end{tabu}}
\end{table}

In both Dist-LeaNet and Amp-PreNet, the neurons of the input layer, hidden layer, and output layer are $N$, $2N$, and $N$, respectively. In Dist-LeaNet, a batch normalization (BN) is employed for the input layer, which normalizes the network input as zero mean and unit variance.
For the hidden layer and output layer of Dist-LeaNet, the linear activation is employed. Then, the Dist-LeaNet is followed by Amp-PreNet with the cascaded mode, i.e., the output of Dist-LeaNet is the input of Amp-PreNet.
Without BN, the leaky rectified unit (LReLU) \cite{III7} and linear activation are adopted for the hidden layer and output layer of Amp-PreNet, respectively.

With the estimated uplink CSI $\widetilde{\bf{g}}_u$, we extract its amplitude feature (denoted as ${\widetilde {\bf{g}}_{u,{\text{amp}}}}$) according to
\begin{equation}
	\label{equ:ul_amp}
	{\widetilde {\bf{g}}_{u,{\text{amp}}}} = {\left[ {\left| {{{\widetilde g}_{u,1}}} \right|,\left| {{{\widetilde g}_{u,2}}} \right|, \ldots ,\left| {{{\widetilde g}_{u,N}}} \right|} \right]^T}.
\end{equation}
Due to the BS's nonlinear distortion (e.g., the LNA and ADC in BS hardware), ${\widetilde {\bf{g}}_{u,{\text{amp}}}}$ cannot use to map the amplitude of downlink CSI of user-$u$ (i.e., the amplitude of $\textbf{h}_u$). This results in that the methods of CSI prediction and recovery in \cite{II3,I2,I3,I5,I6,II2}, cannot be applied directly. Thus, we develop Dist-LeaNet to learn along the direction of the amplitude of $\textbf{g}_u$ and restrain the nonlinear distortion at the BS, which is expressed as
\begin{equation}
	{\widehat {\bf{g}}_{u,{\text{amp}}}} = {f_{{\text{Dist-Lea}}}}\left( {{{\widetilde {\bf{g}}}_{u,{\text{amp}}}},{\bf{\Theta} _{{\text{Dist-Lea}}}}} \right),
\end{equation}
where ${f_{{\text{Dist-Lea}}}}\left(  \cdot  \right)$ and ${{\bf{\Theta} _{{\text{Dist-Lea}}}}}$ denote the mapping function of distortion suppression and the training parameters of Dist-LeaNet, respectively.

On the basis of the obtained ${\widehat {\bf{g}}_{u,{\text{amp}}}}$, the amplitude of downlink CSI of user-$u$ (i.e., the amplitude of $\textbf{h}_u$) can be mapped. Thus, based on the strong amplitude correlation, we construct Amp-PreNet to predict the amplitude feature of downlink CSI ${\widehat {\bf{h}}_{u,{\text{amp}}}}$, which can be expressed as
\begin{equation}
	{\widehat {\bf{h}}_{u,{\text{amp}}}} = {f_{{\text{Amp-Pre}}}}\left( {{{\widehat {\bf{g}}}_{u,{\text{amp}}}},{\bf{\Theta} _{{\text{Amp-Pre}}}}} \right),
\end{equation}
where ${f_{{\text{Amp-Pre}}}}\left(  \cdot  \right)$ and ${{\bf{\Theta} _{{\text{Amp-Pre}}}}}$ denote the mapping function of amplitude prediction and the training parameters of Amp-PreNet, respectively.

\subsubsection{Training and Deployment}
\label{section:3.2.2}

The training sets are acquired by simulation, and a significant amount of data samples are collected to train Dist-LeaNet and Amp-PreNet. Specifically, these data samples are generated as follows.

Amplitude correlated channels are generated by MATLAB 5G Toolbox, which is subject to specifications of the Clustered-Delay-Line (CDL) channel model in 3GPP TR 38.901 \cite{III9}. Similar to the setting in \cite{III10}, the frequency-independent parameters (e.g., the angle of departure (AoD)) are fixed, while varying the complex gain of each path between the uplink and downlink channels.
The AoD of downlink CSI is approximately the same as the angle of arrival (AoA) of the uplink CSI in a short time slot \cite{III10}, showing a relatively strong amplitude correlation in the angle domain. The amplitude attenuation of clusters also reflects the amplitude reciprocity \cite{I2}, due to the similar geographical environment in a short time slot.
Thus, ${{\bf{g}}_u}$ and ${{\bf{h}}_u}$ are obtained by transforming the generated uplink and downlink channels to the angular domain, respectively \cite{III10}.

To train Dist-LeaNet and Amp-PreNet, we use the amplitude of ${{\bf{g}}_u}$ and ${{\bf{h}}_u}$ as network labels, respectively, with the joint training method. The optimization goal of Dist-LeaNet is to minimize the mean squared error (MSE) between ${\widehat {\bf{g}}_{u,{\text{amp}}}}$ and ${{\bf{g}}_{u,{\text{amp}}}}$, which is derived as
\begin{equation}
	\mathop {\min }\limits_{{\bf{\Theta} _{{\text{Dist-Lea}}}}} E\left[ {{{\left\| {{f_{{\text{Dist-Lea}}}}\left( {{{\widetilde {\bf{g}}}_{u,{\text{amp}}}},{\bf{\Theta} _{{\text{Dist-Lea}}}}} \right) - {{\bf{g}}_{u,{\text{amp}}}}} \right\|}^2}} \right].
\end{equation}
Similarly, the Amp-PreNet minimizes the MSE of the downlink CSI amplitude, i.e., $E\left[ {{{\left\| {{{\widehat {\bf{h}}}_{u,{\text{amp}}}} - {{\bf{h}}_{u,{\text{amp}}}}} \right\|}^2}} \right]$, which is further expressed by
\begin{equation}
	\mathop {\min }\limits_{{\bf{\Theta} _{{\text{Amp-Pre}}}}} E\left[ {{{\left\| {{f_{{\text{Amp-Pre}}}}\left( {{{\widehat {\bf{g}}}_{u,{\text{amp}}}},{\bf{\Theta} _{{\text{Amp-Pre}}}}} \right) - {{\bf{h}}_{u,{\text{amp}}}}} \right\|}^2}} \right].
\end{equation}
We perform the joint training once for both Dist-LeaNet and Amp-PreNet, and save the trained network parameters for testing.

By using the Dist-LeaNet, the high precision uplink CSI amplitude ${\widehat {\bf{g}}_{u,{\text{amp}}}}$ is obtained. Then, ${\widehat {\bf{g}}_{u,{\text{amp}}}}$ is used to predict the downlink CSI amplitude ${\widehat {\bf{h}}_{u,{\text{amp}}}}$ in the Amp-PreNet. We consider the distortion before utilizing the channel reciprocity to accomplish amplitude prediction of downlink CSI. The proposed scheme demonstrates a better prediction accuracy and reduces the impact of time delay effectively in a practical scenario.

\section{Experiment results}
\label{section:4}
In this section, we provide numerical results of the proposed scheme. Definitions and basic parameters involved in simulation are first given in Section \ref{section:4.1}. Subsequently, to verify the effectiveness of the proposed scheme, the normalized mean squared error (NMSE) of the predicted downlink CSI amplitude is given in Section \ref{section:4.2}. Finally, computational complexity and online running time comparison analysis are shown in Section \ref{section:4.3}.

\subsection{Parameters Setting}
\label{section:4.1}
\begin{figure}[t]
	\centering
	\includegraphics[width=4in]{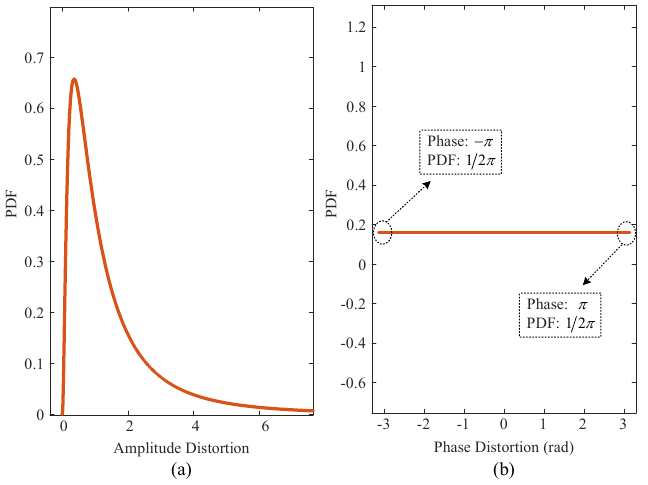}
	\caption{(a) The PDF of the amplitude distortion at the receiver, where $\delta _{{\text{r,BS}}}^2 = 1$.
		(b) The PDF of the phase distortion at the receiver, where ${\theta _{{\text{r,BS}}}} = \pi $.\label{fig_fdis}}
\end{figure}

Definitions involved in simulations are given as follows. The equivalent signal-to-noise ratio (SNR) and NMSE are defined as similar to \cite{II3}.
During the experiments, $v = 300$ km/h, $\delta _{{\text{r,BS}}}^2 = 1$, and ${\theta _{{\text{r,BS}}}} = \pi $ are considered, respectively.
The probability density functions (PDF) of the amplitude and phase of receiver distortion are shown in Fig.~\ref{fig_fdis}.
Following the setting in \cite{I2}, we set the uplink frequency to 5.1 GHz and the downlink frequency to 5.3 GHz. Thus, according to ${f_m} = {{v{f_c}} \mathord{\left/{\vphantom {{v{f_c}} c}} \right.\kern-\nulldelimiterspace} c}$ \cite{RPAbook} with ${f_m}$, ${f_c}$, and ${c}$ being the maximum doppler shift, the carrier frequency, and the speed of light, respectively, the maximum doppler shift for the uplink CSI and downlink CSI are 1418 Hz and 1473 Hz, respectively.
The complex-valued Zadoff-Chu (ZC) sequence \cite{ZC} is employed as the pilot for uplink channel estimation with least squares (LS) criterion in the simulation.
For Dist-LeaNet and Amp-PreNet, their training and testing data-sets are generated according to (\ref{equ:ul_amp}).
The sample numbers of training set, validation set, and testing set are 30,000, 5,000, and 15,000, respectively. In this paper, the NMSE performance of the proposed scheme is compared with those of \cite{II3} and \cite{I2}. In addition, to verify the effectiveness of Dist-LeaNet, the proposed scheme without Dist-LeaNet, denoted as ``Proposed (without Dist-LeaNet)", is also simulated.
It is worth noting that inspired by signal detection \cite{R1-2,R1-3}, the detection-based amplitude prediction is an interesting topic. However, this is beyond the scope of this paper and prompts us to conduct exploratory research in the future.

\subsection{NMSE Performance}
\label{section:4.2}
\begin{figure}[t]
	\centering
	\includegraphics[width=4in]{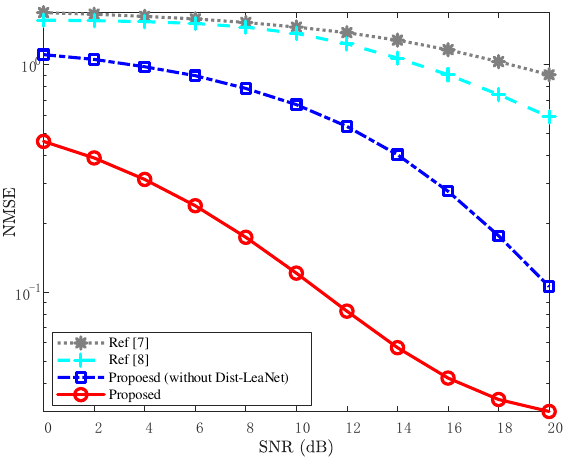}
	\caption{NMSE of downlink CSI amplitude versus SNR, where $N = 128$.\label{fig2}}
\end{figure}

To validate the effectiveness of amplitude prediction, NMSE curves of the recovered downlink CSI amplitude are plotted in Fig.~\ref{fig2}, where $N = 128$ is considered.
From Fig.~\ref{fig2}, it can be observed that the NMSE of ``Proposed" is smaller than those of ``Ref \cite{II3}", ``Ref \cite{I2}", and ``Proposed (without Dist-LeaNet)", showing the effectiveness of the proposed scheme in recovering the downlink CSI amplitude.
Specifically, the NMSE of ``Proposed" is smaller than that of ``Proposed (without Dist-LeaNet)", which confirms that Dist-LeaNet plays an essential role for the proposed scheme in distortion suppression and reciprocity calibration.
In addition, for each given SNR, the NMSE of ``Proposed (without Dist-LeaNet)" is lower than those of ``Ref \cite{II3}" and ``Ref \cite{I2}", which indicates the effectiveness of Amp-PreNet in CSI prediction.
Overall, the proposed scheme proves to be advantageous in improving the NMSE performance in various SNR scenarios.

\begin{figure}[t]
	\centering
	\includegraphics[width=4in]{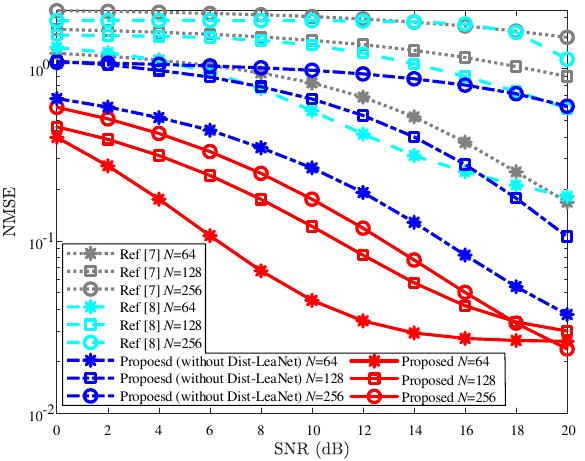}
	\caption{NMSE of downlink CSI amplitude versus SNR under different $N$.\label{fig4}}
\end{figure}

To verify the NMSE performance against the impact of $N$, NMSE curves of ``Ref \cite{II3}", ``Ref \cite{I2}", ``Proposed (without Dist-LeaNet)", and ``Proposed" are plotted in Fig.~\ref{fig4}, where $N = 64$, $N = 128$, and $N = 256$ are considered.
For each given $N$, the NMSE of downlink CSI amplitude of ``Proposed" is smaller than those of ``Ref \cite{II3}", ``Ref \cite{I2}", and ``Proposed (without Dist-LeaNet)".
As the increase of $N$ (i.e., the number of antennas increases), the NMSE increases due to the more nonlinear distortion introduced on antennas at the same SNR.
Thus, compared with ``Ref \cite{II3}", ``Ref \cite{I2}", and ``Proposed (without Dist-LeaNet)", the proposed Dist-LeaNet restrains the distortion and Amp-PreNet predicts the CSI amplitude effectively against varying $N$.

\subsection{Computational Complexity and Online Running Time}
\label{section:4.3}
In this subsection, the computational complexity and online running time of ``Ref \cite{II3}", ``Ref \cite{I2}", and ``Proposed" are presented and analyzed as follows.
\subsubsection{Computational Complexity Analysis}
\label{section:4.3.1}
\begin{figure}[t]
	\centering
	\includegraphics[width=5in]{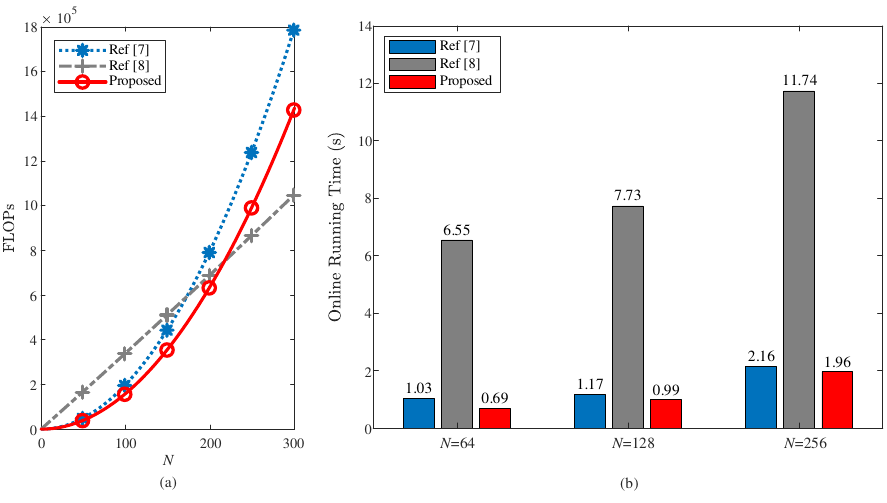}
	\caption{(a) The FLOPs number versus $N$.
		(b) Online running time comparison of ``Ref \cite{II3}", ``Ref \cite{I2}" and ``Proposed" for $10^5$ experiments.\label{fig3}}
\end{figure}

The number of floating-point operations (FLOPs) is considered as the metric of computational complexity, which can be used to describe the NN complexity \cite{III10}. According to \cite{III10}, the FLOPs of ``Ref \cite{II3}", ``Ref \cite{I2}" and ``Proposed" are $20{N^2} - 6N$, ${{{N^2}} \mathord{\left/{\vphantom {{{N^2}} 2}} \right.\kern-\nulldelimiterspace} 2} + {{26841N} \mathord{\left/{\vphantom {{26841N} 8}} \right.\kern-\nulldelimiterspace} 8}$, and $16{N^2} - 6N$, respectively.
The comparison and case details of computational complexity are given in Table~\ref{table_II} and Fig.~\ref{fig3} (a). For $N < 217$, the proposed scheme demonstrates the lowest computational complexity. When $N \geqslant 217$, the FLOPs number of ``Ref \cite{I2}" is lower than those of ``Proposed" and ``Ref \cite{II3}". Nevertheless, the proposed scheme improves the prediction performance of downlink CSI amplitude greatly at the expense of a tolerable computational complexity.

\begin{table}[]
	
	\renewcommand{\arraystretch}{1.2}
	\caption{Analysis of Computational Complexity.}
	\label{table_II}
	\centering
	\scalebox{0.8}{
		\begin{tabu}{@{}c|c|c|c|c@{}}
			\tabucline[0.8pt]{-}
			Method         & Complexity & Case1 $\left( {N = 64} \right)$ & Case2 $\left( {N = 128} \right)$ & Case3 $\left( {N = 256} \right)$ \\ \tabucline[0.6pt]{-}
			Ref \cite{II3} & $20{N^2} - 6N$ & 81,536 & 326,912 & 1,309,180\\ \hline
			Ref \cite{I2}  & ${{{N^2}} \mathord{\left/{\vphantom {{{N^2}} 2}} \right.\kern-\nulldelimiterspace} 2} + {{26841N} \mathord{\left/{\vphantom {{26841N} 8}} \right.\kern-\nulldelimiterspace} 8}$ & 216,776 & 437,648 & 891,680\\ \hline
			Proposed       & $16{N^2} - 6N$ & 65,152 & 261,376 & 1,047,040\\ \tabucline[0.8pt]{-}
	\end{tabu}}
\end{table}

\subsubsection{Online Running Time}
\label{section:4.3.2}
The comparison of online running time is given in Fig.~\ref{fig3} (b), where $N = 64$, $N = 128$, and $N = 256$ are considered. For a fair comparison, $10^5$ experiments of online running are conducted for ``Ref \cite{II3}", ``Ref \cite{I2}" and ``Proposed" on the same computer. From Fig.~\ref{fig3} (b), for each given $N$, the online running time of ``Proposed" is shorter than those of ``Ref \cite{II3}" and ``Ref \cite{I2}".
This reflects that the proposed scheme reduces the transmission and processing delay effectively, due to the application of prediction method and lightweight network architecture.
Additionally, to demonstrate that the proposed scheme can prevent the downlink CSI outdated, the online running time is compared with the coherence time of downlink CSI.
Considering that the correlation coefficient of the channel at any two time points within the coherent time is not less than 0.5 \cite{R2-4}, the maximum Doppler frequency shift ${f_m}$ is used to measure the coherence time of the channel. According to $M = {9 \mathord{\left/{\vphantom {9 {\left( {16\pi {f_m}} \right)}}} \right.\kern-\nulldelimiterspace} {\left( {16\pi {f_m}} \right)}}$ \cite{RPAbook}, the coherence time $M$ of the downlink CSI is 0.122 ms. However, when $N = 64$, $N = 128$, and $N = 256$ are considered, the online running times for each experiment of the proposed scheme are 0.0069 ms, 0.0099 ms, and 0.0196 ms, respectively, which are smaller than the coherence time of downlink CSI. This indicates that although there is a certain delay, the proposed scheme can effectively prevent the predicted downlink CSI amplitude from becoming outdated.

\section{Conclusion}
\label{section:5}
This paper presents an amplitude prediction scheme from uplink to downlink CSI against receiver distortion in FDD systems. By using a lightweight and dedicated Dist-LeaNet, the amplitude feature of the uplink wireless propagation channel is obtained after distortion suppression and reciprocity calibration. Then, with the uplink CSI, the downlink CSI amplitude is predicted by a lightweight Amp-PreNet. Experiments show that, compared with methods that don't consider the distortion in communication systems, the proposed scheme is more practical and achieves a better prediction accuracy on NMSE performance of the downlink CSI amplitude. This idea of considering and handling distortion has reference significance for both UE and BS. In our future work, we will conduct exploratory research on detection-based amplitude prediction methods.

\section*{Acknowledgements}

This work is supported in part by the Sichuan Science and Technology Program (Grant No. 2023YFG0316), the Industry-University Research Innovation Fund of China University (Grant No. 2021ITA10016), the Key Scientific Research Fund of Xihua University (Grant No. Z1320929), and the Special Funds of Industry Development of Sichuan Province (Grant No. zyf-2018-056).





\bibliographystyle{elsarticle-num-names} 
\bibliography{REVISED_Manuscript}

\end{document}